\documentclass[12pt]{article}
\usepackage{epsfig,axodraw}
\def \be {\begin{equation}}
\def \ee {\end{equation}}
\def \bea {\begin{eqnarray}}
\def \eea {\end{eqnarray}}
\def \nn {\nonumber}

\def \a {\alpha}
\def \b {\beta}
\def \g {\gamma}

\def \d {\delta}

\def \m {\mu}
\def \n {\nu}
\def \k {\kappa}
\def \lam {\lambda}

\def \s {\sigma}
\def \r {\rho}
\def \o {\omega}

\def \th {\theta}
\def \Th {\Theta}

\def \t {\tau}
\def \dag {\dagger}
\def \p {\partial}

\def\bd{\begin{document}}
\def\ed{\end{document}}
\def\nn{\nonumber}
\def\bea{\begin{eqnarray}}
\def\eea{\end{eqnarray}}
\let\bm=\bibitem
\let\la=\label

\def\N{{\cal N}}
\def\sst{\scriptscriptstyle}
\def\thetabar{\bar\theta}
\def\Tr{{\rm Tr}}
\def\one{\mbox{1 \kern-.59em {\rm l}}}

\def\a{\alpha}      \def\da{{\dot\alpha}}
\def\b{\beta}       \def\db{{\dot\beta}}
\def\c{\gamma}  \def\C{\Gamma}  \def\cdt{\dot\gamma}
\def\d{\delta}  \def\D{\Delta}  \def\ddt{\dot\delta}
\def\e{\epsilon}        \def\vare{\varepsilon}
\def\f{\phi}    \def\F{\Phi}    \def\vvf{\f}
\def\h{\eta}
\def\k{\kappa}
\def\l{\lambda} \def\L{\Lambda}
\def\m{\mu} \def\n{\nu}
\def\o{\omega}
\def\P{\Pi}
\def\r{\rho}
\def\s{\sigma}  \def\S{\Sigma}
\def\t{\tau}
\def\th{\theta} \def\Th{\Theta} \def\vth{\vartheta}
\def\X{\Xeta}
\def\z{\zeta}
\def\w{\wedge}
\def\u{\underline}

\def\cA{{\cal A}} \def\cB{{\cal B}} \def\cC{{\cal C}}
\def\cD{{\cal D}} \def\cE{{\cal E}} \def\cF{{\cal F}}
\def\cG{{\cal G}} \def\cH{{\cal H}} \def\cI{{\cal I}}
\def\cJ{{\cal J}} \def\cK{{\cal K}} \def\cL{{\cal L}}
\def\cM{{\cal M}} \def\cN{{\cal N}} \def\cO{{\cal O}}
\def\cP{{\cal P}} \def\cQ{{\cal Q}} \def\cR{{\cal R}}
\def\cS{{\cal S}} \def\cT{{\cal T}} \def\cU{{\cal U}}
\def\cV{{\cal V}} \def\cW{{\cal W}} \def\cX{{\cal X}}
\def\cY{{\cal Y}} \def\cZ{{\cal Z}}

\def\ua{\underline{\alpha}}
\def\ub{\underline{\phantom{\alpha}}\!\!\!\beta}
\def\uc{\underline{\phantom{\alpha}}\!\!\!\gamma}
\def\um{\underline{\mu}}
\def\ud{\underline\delta}
\def\ue{\underline\epsilon}
\def\una{\underline a}\def\unA{\underline A}
\def\unb{\underline b}\def\unB{\underline B}
\def\unc{\underline c}\def\unC{\underline C}
\def\und{\underline d}\def\unD{\underline D}
\def\une{\underline e}\def\unE{\underline E}
\def\unf{\underline{\phantom{e}}\!\!\!\! f}\def\unF{\underline F}
\def\unm{\underline m}\def\unM{\underline M}
\def\unn{\underline n}\def\unN{\underline N}
\def\unp{\underline{\phantom{a}}\!\!\! p}\def\unP{\underline P}
\def\unq{\underline{\phantom{a}}\!\!\! q}
\def\unQ{\underline{\phantom{A}}\!\!\!\! Q}
\def\unH{\underline{H}}

\def\As {{A \hspace{-6.4pt} \slash}\;}
\def\bs {{b \hspace{-6.4pt} \slash}\;}
\def\Ds {{D \hspace{-6.4pt} \slash}\;}
\def\ds {{\del \hspace{-6.4pt} \slash}\;}
\def\ss {{\s \hspace{-6.4pt} \slash}\;}
\def\ks {{ k \hspace{-6.4pt} \slash}\;}
\def\ps {{p \hspace{-6.4pt} \slash}\;}
\def\pas {{{p_1} \hspace{-6.4pt} \slash}\;}
\def\pbs {{{p_2} \hspace{-6.4pt} \slash}\;}

\def\Fh{\hat{F}}
\def\Vh{\hat{V}}
\def\Xh{\hat{X}}
\def\ah{\hat{a}}
\def\xh{\hat{x}}
\def\yh{\hat{y}}
\def\ph{\hat{p}}
\def\xih{\hat{\xi}}

\def\psit{\tilde{\psi}}
\def\Psit{\tilde{\Psi}}
\def\tht{\tilde{\th}}

\def\At{\tilde{A}}
\def\Qt{\tilde{Q}}
\def\Rt{\tilde{R}}
\def\Nt{\tilde{N}}

\def\at{\tilde{a}}
\def\st{\tilde{s}}
\def\ft{\tilde{f}}
\def\pt{\tilde{p}}
\def\qt{\tilde{q}}
\def\vt{\tilde{v}}
\def\nt{\tilde{n}}

\def\delb{\bar{\partial}}
\def\bz{\bar{z}}
\def\bD{\bar{D}}
\def\bB{\bar{B}}

\def\bk{{\bf k}}
\def\bl{{\bf l}}
\def\bp{{\bf p}}
\def\bq{{\bf q}}
\def\br{{\bf r}}
\def\bx{{\bf x}}
\def\by{{\bf y}}
\def\bR{{\bf R}}
\def\bV{{\bf V}}

\def\d{\delta}\def\D{\Delta}\def\ddt{\dot\delta}

\def\p{\partial} \def\del{\partial}
\def\xx{\times}
\def\uno{\mbox{1 \kern-.59em {\rm l}}}

\def\trp{^{\top}}
\def\inv{^{-1}}
\def\dag{{^{\dagger}}}
\def\pr{^{\prime}}

\def\rar{\rightarrow}
\def\lar{\leftarrow}
\def\lrar{\leftrightarrow}

\def\bc{\begin{center}}
\def\ec{\end{center}}
\def\bp{\begin{picture}}
\def\ep{\end{picture}}

\def\bps{ eq.~(\ref{halfbps})}

\begin{document}
\title{\bf Supersymmetric Wilson Loops in ${\cal N}=6$ Super Chern-Simons-matter theory}

\author{Bin Chen\\
Department of Physics,\\
and State Key Laboratory of Nuclear Physics and Technology,\\
Peking University,\\
Beijing 100871, P.R. China\\
E-mail: bchen01@pku.edu.cn\\ \\
Jun-Bao Wu\\International School for Advanced Studies (SISSA), \\
via Beirut 2-4, I-34014 Trieste, Italy\\
and INFN, Trieste section, Via Valerio 2, I - 34127 Trieste, Italy\\
E-mail: wujunbao@sissa.it}

 \maketitle

\begin{abstract}
We study supersymmetric Wilson loop operators in ABJM theory from
both sides of the $AdS_4/CFT_3$ correspondence. We first construct
some supersymmetric Wilson loops. The perturbative computations are
performed in the field theory side at the first two orders. A
fundamental string solution ending on a circular loop is also
studied.
\end{abstract}

\section{Introduction}

Four-dimensional ${\cal N}=4$ super Yang-Mills theory is one of the
most remarkable field theories. Among other things, this theory
appears as low energy effective theory of multi D3-branes. This
leads to the famous $AdS_5/CFT_4$ correspondence \cite{Mal97,
Gubser:1998bc, Witten:1998qj, Aharony:1999ti}. Besides the local
operators, supersymmetric Wilson loops play an important role in
this correspondence. The vacuum expectation value (VEV) of half-BPS
straight Wilson lines equals to unit due to the global
supersymmetries they preserve. But the vacuum expectation value of
half-BPS circular Wilson loop depends on the coupling constant in a
nontrivial way. The computation of this VEV can be reduced to a
computation via a Gaussian matrix model. Evidence of this reduction
was found first by direct perturbative calculations \cite{Zarembo}:
in the Feynman gauge, the contributions from the diagrams which are
not ladder/rainbow diagrams cancel each other. Later, this reduction
was obtained from the conformal transformation linking the circular
loop and the straight line \cite{Drukker01}. Recently this reduction
was obtained through localization technique \cite{Pestun:2007rz}.
For the Wilson loop in the fundamental representation of the gauge
group, the VEV computed via this matrix model has the same large
$N$, large $\lam$ behavior as the one from the string theory dual
description in terms of fundamental strings in $AdS_5$ ending on
this loop \cite{SJR, Maldacena98, Berenstein:1998ij,
Drukker:1999zq}. Similar conclusion was found for Wilson loops in
symmetric (anti-symmetric) representation, in which case a better
dual description is in terms of  $D3$ ($D5$) branes \cite{Drukker,
Yamaguchi:2006D5, Gomis:2006sb, Gomis:2006im,ChenHe}. Various
aspects of Wilson loops with less supersymmetries were discussed in
\cite{Zarembo2}-\cite{Young:2008ed}.

The low energy effective theory of multi M2-branes and M5-branes is
one of the most important subjects in M-theory. For multi M2-branes
in flat eleven dimensional spacetime, the low energy effective
theory is the IR limit of ${\cal N}=8$ three dimensional super
Yang-Mills theory. The construction of ${\cal N}=8$ three
dimensional conformal field theory for multi-membranes has been a
vexing question for more than ten years. %perplexed People have
%looked for equivalent description of this theory for many years,
%with the hope that the new description may be easier to study.
Recently, based on early works \cite{Harvey}, this question was
addressed \cite{BL, G} with the help of a novel 3-algebra. The
theory based on seemingly the only unitary 3-algebra was found to be
a theory for two M2-branes on an orbifold \cite{Mukhi, LambertTong}.
More interestingly it could be rewritten as a super-Chern-Simons
theory of two gauge groups with opposite level\cite{Raamsdonk}. This
fact, combined with the earlier attempts of constructing
three-dimensional super-conformal field theory as Chern-Simons
theory \cite{Schwarz,GY}, inspired people to search for the
Chern-Simons-matter theory with higher supersymmetry. In \cite{GW},
the one with ${\cal N}=4$ supersymmetry has been constructed.  Very
recently, a ${\cal N}=6$ super Chern-Simons-matter theory was
proposed in \cite{ABJM} to describe  arbitrary number of M2-branes
on the orbifold $C^4/Z_k$ (For some  other discussions on
three-dimension ${\cal N}=4, 5, 6$ super Chern-Simons-matter
theories, see \cite{Hosomichi:2008jd}-\cite{ABJ}.) . In \cite{ABJM},
the dual M/string description of large N limit of three-dimensional
conformal theory was discussed. Depending on different values of the
parameters in the field theory, this so-called ABJM theory can be
dual to M-theory on $AdS_4\times S^7/Z_k$ or IIA theory on
$AdS_4\times CP^3$.

The subject of this paper is to study the supersymmetric Wilson
loops in the ABJM theory. The pure Chern-Simons theory is a
topological theory and the correlation functions
 of Wilson loops in this theory give various knot invariants \cite{Witten89}. The ABJM
theory includes matters and is no more a topological theory. We
construct $1/6$-BPS Wilson loops which involve the matter fields.
Both the dimensional considerations and the supersymmetries lead us
to include the product of two scalar fields in these operators.

As a first step to study these BPS Wilson loops, we perform the
perturbative computations up to the order of $\lambda^2$ in the
planar limit. %It is an interesting problem to perform this
%perturbative computations at higher orders.
One the other hand, we find a fundamental string solution ending on
a circular loop on the boundary. The dual fundamental string
solution here is quite similar to the one in $AdS_5$ case. This
leads us to expect that this $F1$-string solution is half-BPS, and
it should be dual to a half-BPS
Wilson loop which is to be constructed. %We leave the studies of the origin of this
%difference to further works.
%We discuss the possible origins of this mismatch.

The other part of this paper is organized as following: in the next
section, we construct some supersymmetric Wilson loops in ABJM
theory and discuss the supersymmetries they preserve. In section
\ref{ft}, we first perform the perturbative calculations at the
first two orders. In section \ref{string}, we will discuss the
string theory dual of the to-be-constructed half-BPS Wilson loops.
The conclusion and discussions are put in the last section. We also
include some needed Feynman rules in Appendix A.\\

{\bf Note added}

When this project was in progress. we learnt that this subject was
also studied by Nadav Drukker, Jan Plefka and Donovan Young, whose
paper \cite{DPY} has overlap with ours.

\section{Supersymmetric Wilson loops in super Chern-Simons-matter theory}
In this paper, we study supersymmetric Wilson loop operators in
three dimensional ${\cal N}=6$ super Chern-Simons-matter theory
proposed in \cite{ABJM}. The gauge group of this theory is
$U_1(N)\times U_2(N)$. The matters include bifundamental superfields
$A_1, A_2$ and anti-bifundamental superfields $B_1, B_2$. The
Chern-Simons part of the action is:
 \bea
 S_{CS}&=&\frac{k}{4\pi}\int {\rm tr}(A_{(1)}\wedge dA_{(1)}+\frac23
 A_{(1)}^3)
-{\rm tr}(A_{(2)}\wedge
 dA_{(2)}+ \frac23 A_{(2)}^3).
 \eea
Notice that the levels of these two group are opposite. The
superpotential of this theory reads: \bea
 W&=&\frac{4\pi}{k}\Tr(A_1B_1A_2B_2-A_1B_2A_2B_1).
 \eea
We denote $(A_1, A_2, \bar{B}_1, \bar{B}_2)$ as $Y^A, A=1, \cdots,
4$ which is in the fundamental representation of the R-symmetry
group $SU_R(4)$.

The supersymmetry transformation of the gauge fields and the scalar
fields are\footnote{We don't present the supersymmetry
transformations of fermions, which are not essential to our study.}
\cite{Terashima:2008sy}: \bea \d A^{(1)}_\mu&=&-\frac{2\pi}k
\left(Y^A\psi^{B\dag}\g_\m\omega_{AB}+\omega^{AB}\g_\m\psi_A
Y^\dag_B\right),\\
\d A^{(2)}_\m&=&\frac{2\pi}k\left(\psi^{A\dag}Y^B\g_\m\omega_{AB}+\omega^{AB}\g_\m Y^\dag_A \psi_B\right),\\
\d Y^A&=&i\omega^{AB}\psi_B,\\
\d Y^\dag_A&=&i\psi^{B\dag}\omega_{AB}, \eea where $\psi_A$ is the
supersymmetry partner of $Y^A$ and $\omega_{AB}=-\omega_{BA}$ are
the supersymmetry parameters and they are $6$ Majorana spinors in
three dimensional spacetime. Here $\omega^{AB}$ is defined as
\be\omega^{AB}=\frac12\epsilon^{ABCD}\omega_{CD}, \ee and we also
have the following relation: \be
(\omega^{AB})_\alpha=\left((\omega_{AB})_\alpha\right)^\ast. \ee
% The supersymmetry transformation is also studied in \cite{...}.

We will search for  supersymmetric spacelike Wilson loop operators
in the following class:
 \bea && \frac1NTr_{U_1(N)}P\exp\left[i\int d\t
\left(A_\m^{(1)}\dot{x}^\m+i s_A^B(x)Y^AY^\dagger_B\right)\right]\nn\\
&\times&\frac1N Tr_{U_2(N)}P\exp\left[ i\int d\t
\left(A_\m^{(2)}\dot{x}^\m+i
\tilde{s}_B^A(x)Y^\dag_AY^B\right)\right],\label{wl} \eea where
$s_A^B$ and $\tilde{s}_B^A$ are Hermitian matrix, $A_\m^{(1)}=A_{\m
a}^{(1)}(T^a)_i^j$, $A_\m^{(2)}=A_{\m
\tilde{a}}^{(2)}(T^{\tilde{a}})_{\tilde{i}}^{\tilde{j}}$. We have
also omitted the indices of $Y^A$, $Y^\dag_A$ which are
$(Y^A)_i^{\tilde j}$ and $(Y^\dag_A)_{\tilde i}^{j}$. The indices
$i, j$ are the color indices of $U_1(N)$ and the indices $\tilde{i},
\tilde{j}$ are the color indices of $U_2(N)$. The normalization of
the generators of the Lie algebra is chosen as $(T^a)_i^j
(T^a)_k^l=\frac12 \d_k^j\d_i^l, (T^{\tilde a})_{\tilde i}^{\tilde j}
(T^{\tilde a})_{\tilde k}^{\tilde l}=\frac12 \d_{\tilde k}^{\tilde
j} \d_{\tilde i}^{\tilde l}$.

For the above Wilson line operators to be supersymmetric, the
supersymmetry transformation of each factor should vanish by itself.
The transformation of the first factor is proportional to \be
-\frac{2\pi}k(Y^A\psi^{B\dag}\g_\m\omega_{AB}+\omega^{AB}\g_\m\psi_AY^\dag_B)\dot{x}^\m
-\omega^{AC}\psi_CY^\dag_Bs_A^B-Y^A\psi^{C \dag}\omega_{BC}s_A^B.\ee
Then we get the following condition for the operator to be
supersymmetric: \be
-\frac{2\pi}k\dot{x}^\m\g_\m\omega_{AB}-s_A^C\omega_{CB}=0.\label{susy1}\ee
\be
-\frac{2\pi}k\omega^{AB}\g_\m\dot{x}^\m-\omega^{CA}s_C^B=0.\label{susy2}\ee
In fact, eq.~(\ref{susy2}) is complex conjugation of
eq.~(\ref{susy1}). Notice that $\omega_{AB}$ is antisymmetric, we
get from eq.~(\ref{susy1}): \be
-\frac{2\pi}k\dot{x}^\m\g_\m\omega_{BA}-s_A^C\omega_{BC}=0,\ee from
which and the following equation from relabelling the indices in
eq.~(\ref{susy1}) \be
-\frac{2\pi}k\dot{x}^\m\g_\m\omega_{BA}-s_B^C\omega_{CA}=0,\ee we
have \be s_A^C\omega_{BC}=s_B^C\omega_{CA}.\label{relation} \ee
Since the general case is not easy to study,  now we consider the
special case in which $s$ and $\tilde{s}$ are diagonal matrices: \be
s_A^B=s_A\d_A^B,\hspace{5ex} \tilde{s}_A^B=\tilde{s}_A\d_A^B. \ee In
this case, from eq.~(\ref{relation}), we get: \be
(s_A-s_B)\omega_{AB}=0, \ee this is to say that if $s_A\ne s_B$,
$\omega_{AB}=0$.

Now we turn to consider the second factor: \be
\frac1NTr_{U_2(N)}P\exp\left[i\int d\t
\left(A_\m^{(2)}\dot{x}^\m+i\tilde{s}_A Y^\dag_A Y^A \right)\right].
\ee Similarly the supersymmetric condition for this factor gives:
\be \frac{2\pi}{k}\dot{x}^\mu \g_\mu \omega_{AB}-\tilde{s}_B
\omega_{BA}=0,\label{susy3} \ee and its complex conjugation. As
before, we can get\be
(\tilde{s}_A-\tilde{s}_B)\omega_{AB}=0.\label{susy4} \ee From
eqs.~(\ref{susy1}) and ~(\ref{susy3}-\ref{susy4}), we get \be
(s_A-\tilde{s}_A)\omega_{AB}=0. \ee So if $s_A \ne \tilde{s}_A$,
$\omega_{AB}$ must be zero.

Using all of the above results, and also taking the following
condition into account: \be
\omega_{AB}^*=\omega^{AB}=\frac12\epsilon^{ABCD}\omega_{CD},
\label{real}\ee and the $\gamma$-matrices in \cite{Benna:2008zy,
Terashima:2008sy}, we find the $1/6-$BPS Wilson loops which is
defined by
 $s_{1, 2}=\tilde{s}_{1, 2}=-s_{3,
4}=-\tilde{s}_{3, 4}=\pm \frac{2\pi}k|\dot{x}|$. Choose the positive
sign for $s_1$, the Wilson loop is \bea
&&\frac1N Tr_{U_1(N)} P {\rm exp}\left( i\int d\t (A^{(1)}_\mu \dot{x}^\mu+i\frac{2\pi}k (A_i \bar{A_i}-\bar{B_i}B_i) |\dot{\vec{x}}| )\right)\nn\\
\times &&\frac1N Tr_{U_2(N)} P {\rm exp}\left( i\int d\t
(A^{(2)}_\mu \dot{x}^\mu+i\frac{2\pi}k(\bar{A_i}A_i-B_i\bar{B_i})
|\dot{\vec{x}}| )\right).\label{halfbps} \eea Here for straight
Wilson line, the preserved supercharges are global.

If we put the spacelike Wilson line along the $x^2$ direction, the
above discussions give that the unbroken supersymmetries
$(Q^{AB})^\a$ are $(Q^{12})^2$ and $(Q^{34})^1$. They satisfy the
following superalgebra: \be \{(Q^{12})^2, (Q^{34})^1\}=-P_2,
\{(Q^{12})^2, (Q^{12})^2\}=\{(Q^{34})^1, (Q^{34})^1\}=0. \ee Notices
that among three $P_\mu$'s, only $P_2$ appears in the right-hand
side of superalgebra. This is as expected since only translation
along $x^2$ direction is unbroken.

The Wilson loops preserve two supercharges in ${\cal N}=2$ and
${\cal N}=3$ Chern-Simons-matter theory were studied in \cite{GY}.
Our $1/6-$BPS operators are the generalization to ABJM theory of
these operators. It is interesting to see if one can find other BPS
Wilson loops within and/or out of the class eq.~(\ref{wl}) we begin
with.

\section{Perturbative computations \label{ft}}

In this section, we perform the perturbative computations of the
vacuum expectation values of the above $1/6-$BPS Wilson loops in the
large $N$ limit. Then the expansion parameter is $\lam=N/k$. We will
do the computation up to the order of $\lam^2$. We list some needed
Feynman rules in Appendix A. And we choose the Landau gauge and use
the regularization by dimensional reduction as done in
\cite{Chen:1992ee}. In this regularization scheme, we first perform
all tensor contractions in strict three dimensions: \be
\epsilon_{\m\n\rho}\epsilon^{\mu\sigma\lambda}=\d^\sigma_\n\d^\lambda_\rho-\d^\lambda_\n\d^\sigma_\rho,
\d^\m_\m=3, \ee then we regular the divergent scalar integrals by
the dimensional continuation with $D=3-\epsilon$.

We will mainly focus on the circular Wilson loops.
% We'll briefly discuss the $1/6$-BPS Wilson loops
% and straight Wilson lines.
We parameterize the circle as  \be x^\m(\t)=(L\cos\t, L\sin\t,
0),\hspace{3ex} 0\le\tau\le 2\pi, \label{circle}\ee with the $x^3$
direction being the time direction.

%As one can see in the following calculations, there are some
%remarkable differences in some aspects from the perturbative
%calculations of the VEV of half-BPS Wilson loops in ${\cal N}=4$
%super Yang-Mills.

\subsection{The calculations at the order of $\lam$}

 \begin{figure}[ht!]
    \epsfxsize=35mm%
    \hfill\epsfbox{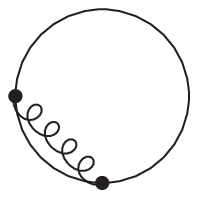}\hfill~\\
    \caption{The only contributing diagram at the order of $\lam$.}
    %\label{figLT}
   \end{figure}

We show in this subsection that the contribution at this order is
zero. Due to the factor $\frac{2\pi}k$ in front of the product of
two scalars, we need only consider the contributions from the gauge
field. There is only one contributing diagram as in Fig.~1. The
contributions from this diagram reads: \be \frac{i^2}N
\int_0^{2\pi}d\t_1\int_0^{\t_1}d\t_2\langle A^{\m_1}(x(\t_1))
A^{\m_2}(x(\t_2))\rangle
\frac{dx_{\m_1}(\t_1)}{d\t_1}\frac{dx_{\m_2}(\t_2)}{d\t_2},\label{eqk}
\ee where the correlator of two gauge field is the free field one.
The momentum space gluon propagator in the Landau gauge is
proportional to \footnote{We don't need to consider the color factor
in the computations at this order.}
\be\frac{\epsilon^{\m\n\lambda}p_\lambda}{p^2}. \ee By Fourier
transformation, we get the position space propagator \be \langle
A^{\m_1}(x(\t_1)) A^{\m_2}(x(\t_2))\rangle \propto
\frac{\epsilon^{\m_1\m_2\lambda}(x_1-x_2)_\lambda}{|x_1-x_2|^3}, \ee
where $x_i=x(\t_i), i=1, 2$. Using this result, and
eq.~(\ref{circle}), one can easily see that eq.~(\ref{eqk}) is equal
to zero.

%The result that eq.~(\ref{eqk}) equals to zero can also be
%obtained from the spacetime parity\footnote{We thank Nadav Drukker
%for discussions on this point.}.
 %(See also Appendix \ref{appendixparity})

\subsection{The calculations at the order of $\lam^2$}

 \begin{figure}[ht!]
    \epsfxsize=50mm%
    \hfill\epsfbox{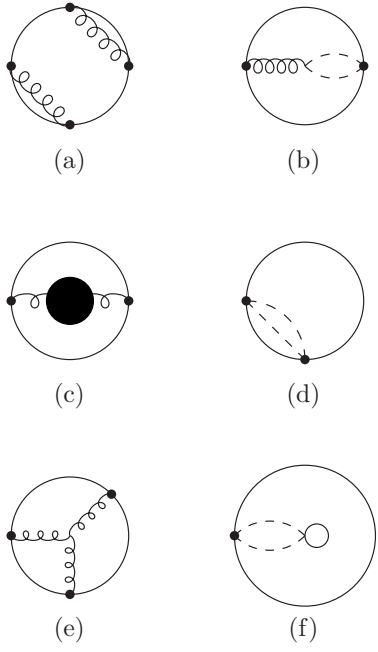}\hfill~\\
    \caption{The contributing diagrams at the order of $\lam^2$.}
    %\label{figLT}
   \end{figure}
%\footnote{We thank Kimyeong Lee for suggesting us
%to consider the Fig.~2.f.}.
We first focus on the contribution due to the expansion of the first
factor. The planar diagrams at the order of $\lam^2$ are listed in
Fig.~2. From the discussions in the previous subsection, we know
that the contribution from Figs.~2.a vanishes. The contribution from
Fig.~2.f also vanishes due to the relative minus sign between $A_i
\bar{A_i}$ and $\bar{B_i}B_i$ in the Wilson loop. The contribution
from Fig~2.b can be computed first in momentum space, then the
result can be transferred to the position space by a Fourier
transformation. The contribution from this diagram is proportional
to \be
\frac{i^2}{Nk^2}\int_0^{2\pi}d\t_1\int_0^{\t_1}d\t_2\dot{x}(\t_1)_\mu|\dot{x}(\t_2)|
\int\frac{d^2p}{(2\pi^3)}e^{\left[ip\cdot
(x(\t_1)-x(\t_2))\right]}\frac{\epsilon^{\mu\nu\lambda}p_\nu}{p^2}\int\frac{d^3q}{(2\pi)^3}\frac{
2 q_{\l}-p_{\l}}{q^2(p-q)^2}.\ee Since $\epsilon^{\mu\n\l}p_\n
p_\l=0$, we need only consider  \be
\int\frac{d^3q}{(2\pi)^3}\frac{q_{\l}}{q^2(p-q)^2}  \ee in the last
integration. From the Lorentz structure we know that this must
proportional to $p_\l$, so this will not contribute, either. We can
check the above claim through direction computations using
regularization by dimensional reduction in \cite{Chen:1992ee}: \bea
&& \int\frac{d^Dq}{(2\pi)^D}\frac{q_{\l}}{q^2(p-q)^2}\\
&=&\int\frac{d^Dq}{(2\pi)^D}\int_0^1 dx\frac{q_{\l}}{[(1-x)q^2+x(p-q)^2]^2}\\
&=&\int_0^1dx\frac{d^Dq}{(2\pi)^D}\frac{q_\l}{[(q-xp)^2+(x-x^2)p^2]^2}
\eea After the shift $q\to q+xp$, we get \be
\int_0^1dx\frac{d^Dq}{(2\pi)^D}\frac{q_\l+xp_\l}{[q^2+(x-x^2)p^2]^2}\ee
the first term vanish and the second term is proportional to $p_\l$,
this confirms our claim.

 \begin{figure}[ht!]
    \epsfxsize=50mm%
    \hfill\epsfbox{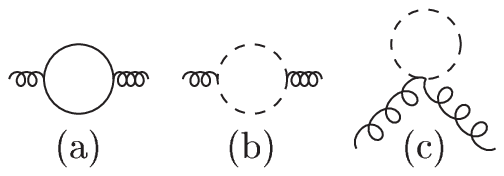}\hfill~\\
    \caption{One-matter-loop correction to the gluon propagator.}
    %\label{figLT}
   \end{figure}

Now we turn to the contributions from Fig.~2.c-2.d. The
contributions from Figs.~2.c are from the one-loop corrections to
gluon propagator, and the contribution from Fig.~2.d includes two
scalar propagators. The combination of these two parts has a quite
remarkable structure, which is the same as the sum of the gluon
propagator in the Landau gauge and the scalar propagator in the four
dimensional ${\cal N}=4$ super Yang-Mills theory.

The sum of the contributions from these diagrams is equal to: \bea
&&\frac{i^2}{N}
\int_0^{2\pi}d\t_1\int_0^{\t_1}d\t_2\left(\dot{x}(\t_1)^\mu
\dot{x}(\t_2)^\nu\langle A^\mu(x(\t_1))_i^j A^\n(x(\t_2))_j^k\rangle_{ol}-\left(\frac{2\pi}k\right)^2\right.\nn\\
&\times&\left.|\dot{x}(\t_1)||\dot{x}(\t_2)|
\langle((A_i\bar{A}_i-\bar{B}_iB_i)(x(\t_1)))_i^j((A_i\bar{A}_i-\bar{B}_iB_i)(x(\t_2)))_j^k\rangle\right),%\nn\\
\label{matrix} \eea where the subscript {\it 'ol'} means one-loop.
The contributions from the gluon loop and the ghost loop cancel each
other \cite{Chen:1992ee}, noticing that we are using the
regularization by dimensional reduction. The contributions from
one-matter-loop are from Figs.~3.a-3.c. Both the one-fermion-loop
and the one-scalar-loop correction to the gluon self-energy are
\cite{Chen:1992ee} (see also \cite{GY})
\be\Pi_{\m\n}(p)=-\frac{N\d^k_i}{64}\left(p\eta_{\m\n}-\frac{p_\m
p_\n}{p}\right). \ee Taking the two gluon propagators into account,
we get:
\bea && \left(\frac{4\pi}{k}\right)^2 \frac{\epsilon^{\m\m_1\m_2}p_{\m_2}}{p^2}\Pi_{\m_1\n_1}(p)\frac{\epsilon^{\n\n_1\n_2}p_{\n_2}}{p^2} \nn\\
&=&-\frac{N\pi^2}{4k^2}\d^k_i\frac{p^2\eta^{\m\n}-p^\m
p^\n}{p^3}.\eea In the ABJM theory, we have $4N$ scalars and $4N$
fermions, so totally we get: \be -\frac{2\pi^2
N^2}{k^2}\d^{ab}\frac{p^2\eta^{\m\n}-p^\m p^\n}{p^3}. \label{propa}
\ee The second term of the above propagator will not contribute to
the final answer due to the gauge invariance. In fact, the Fourier
transformation of this term is proportional to \be
\p_\m^1\p_\n^2\log|x_1-x_2|, \ee where $x_i=x(\tau_i),
\p_\mu^i=\frac{\p}{\p x^\m_i}, i=1, 2$. So its contribution to
eq.~(\ref{matrix}) is proportional to \be
\int_0^{2\pi}d\t_1\int_0^{\t_1}d\t_2\dot{x}^{\m}(\t_1)\dot{x}^\n(\t_2)\p_\m^1\p_\n^2\log|x_1-x_2|,
\ee now we regularize the above integration as: \be
\int_0^{2\pi}d\t_1\dot{x}^{\m}(\t_1)\int_0^{\t(\tilde\epsilon)}d\t_2\frac{d}{d\t_2}(\p_\m^1\log|x_1-x_2|),
\ee where $\t(\tilde\epsilon)=\t_1-\tilde\epsilon$ and
$\tilde\epsilon$ is a regularization parameter. Then we get \be
\int_0^{2\pi}d\t_1\dot{x}^\m(\t_1)(\p_\m^1\log|x_1-x(\t(\tilde\epsilon))|-\p_\m^1\log|x_1-x(0)|)
\ee Using $\p_\mu^1\log|x_1-y|\propto x_{\mu}(\tau_1)$ and
$x(\tau_1)\cdot\dot{x}(\t_1)=0$, we get that eq. (2) is zero even
before we take $\tilde\epsilon\to 0$. So this term doesn't
contribute as claimed.

By performing a Fourier transformation of the first term in
eq.~(\ref{propa}) to position space, we get \be
\int\frac{d^3p}{(2\pi)^3}e^{i p\cdot (x-y)}\left(-\frac{2\pi^2
N^2}{k^2}\d^k_i\frac{1}{p^3}\right)=\frac{N^2\eta^{\m\n}\d^{ab}}{k^2(x-y)^2}.
\ee
%This is the full one-matter-loop contributions to gluon propagator.
So the contribution to \be i^2 \dot{x}_\m \dot{y}_\n \langle
(A^\mu(x))_i^j (A^\n (y))_j^k \rangle \ee at this order equals to
\be -\frac{N^2}{k^2}\frac{\dot{x}\cdot\dot{y}}{(x-y)^2}\d^k_i,
\label{eqvector}\ee where $x\equiv x_1, y\equiv x_2$. Now we turn to
discuss the contribution due to the product of two scalar
propagators which appears in Fig.~2.d. Using the position space
 propagator of (anti-)bifundamental scalars   \be \frac{1}{4\pi}\frac1{|x-y|}\d^l_i\d^{\tilde
j}_{\tilde k}, \ee we get that \bea
-\left(i\frac{2\pi}k\right)^2|\dot{x}||\dot{y}|\langle(A_i\bar{A}_i-B_i\bar{B}_i)_i^j(x)
(A_j\bar{A}_j-B_j\bar{B}_j)_j^k(y)\rangle=
\frac{N^2}{k^2}\frac{|\dot{x}||\dot{y}|}{|x-y|^2}\d_i^k.\label{eqscalar}
\eea

The sum of eqs.~(\ref{eqvector}) and (\ref{eqscalar}) equals to \be
-\frac{N^2}{k^2}\frac{\dot{x}\cdot\dot{y}-|\dot{x}||\dot{y}|}{|x-y|^2}\d_i^k.
\ee When $x(\tau)$ is among the circle in eq.~(\ref{circle}), the
above result equals to \be \frac{N^2}{2k^2}\d_i^k \ee which is
independent of the positions.

From this we get that eq.~(\ref{matrix}) is equal to \be
\frac{N^2}{2k^2}\int_0^{2\pi}d\t_1\int_0^{\t_1}d\t_2=\frac{\pi^2N^2}{k^2}.
\ee The second factor of the Wilson loop gives the same
contribution. Then the total contribution is
$\frac{2\pi^2N^2}{k^2}$.
%When $x(\tau)$ is along a straight line, the above result vanishes.

Now at the order of $\lam^2$, we are left with the contribution from
Fig.~2.e. %which is out of the class defined in the previous paragraph.
This diagram comes from the pure Chern-Simons theory and was studied
in details in \cite{Guadagnini:1989am}. The result from each factor
in the Wilson line is $-\pi^2N^2/(6k^2)$. So totally the
contributions at the order of $\lambda^2$ is
\begin{equation}
2\pi^2\lambda^2-2\cdot\frac{\pi^2\lambda^2}6=\frac{5}{3}\pi^2\lambda^2.
\end{equation}

\section{Holographic description\label{string}}

Depending on the values of $N$ and $k$, the gravity dual of the ABJM
theory can be M-theory on $AdS_4\times S^7/Z_k$ or type IIA theory
on $AdS_4\times CP^3$ with two-form and four-form fluxes. In this
section we will focus on the type IIA case which corresponds to the
situation when $k\ll N\ll k^5$. As the four dimensional super
Yang-Mills case, the Wilson loop in the fundamental representation
of the gauge group should be dual to a fundamental string in this
background while the better description for the Wilson loop in
higher dimensional representation should be higher dimensional probe
D-branes or bubbling geometry.
%We only consider the case when the backreaction of the probe string or
%brane can be neglected.

In this section, we find a fundamental string solution ending on a
circle on the boundary.
%\subsection{Fundamental string description}
The dual IIA string theory background is (We take
$\alpha^\prime=1$):
\bea ds^2_{string}&=&\frac{R^3}k(\frac14 ds^2_{AdS_4}+ds^2_{CP^3}),\\
e^{2\phi}&=&\frac{R^3}{k^3},\\
F_4&=&\frac38R^3\epsilon_{AdS_4},\\
F_2&=&kJ, \eea where $\epsilon_{AdS_4}$ is the volume form of unit
$AdS_4$ and $J$ is the Kahler form of $CP^3$, and $R$ satisfies \be
R^3=2^{5/2}\pi\sqrt{Nk}. \ee

Using the Poincare coordinates in the $AdS_4$ part, the metric can
be written as \be
ds^2=\frac{R^3}{4k}\frac1{y^2}(dy^2+dx_1^2+dx_2^2+dx_3^2)+\frac{R^3}{k}ds^2_{CP^3},
\ee where $y=0$ is the conformal boundary of $AdS_4$. We put the
circular Wilson loop at $x_1^2+x_2^2=R^2$ and $x_3=0$ on the
boundary. As in the $AdS_5$ case \cite{SJR, Maldacena98}, the
worldsheet of this fundamental string should be completely embedded
in $AdS_4$ ending on the Wilson loop \cite{ABJM}. The Nambu-Goto
action of the fundamental string is \be S_{NG}=\frac{1}{2\pi}\int
d\sigma d\t \sqrt{{\rm det}\tilde{g}}, \ee where $\tilde{g}$ is the
induced metric on the worldsheet. The fundamental string solution we
are looking for is the same as the $AdS_5$ case
\cite{Berenstein:1998ij, Drukker:1999zq}: \be x_1^2+x_2^2+y^2=R^2.
\ee

The total action of this string is equal to \be
-\frac{R^3}{4k}=-\pi\sqrt{2\lambda}. \ee

Since this fundamental string solution is quite similar to the one
in IIB theory on $AdS_5\times S^5$, we expect that this solution is
half-BPS. It should be very interesting to construct the dual
half-BPS Wilson loop in the field theory side.
%The difference may due
%to some non-perturbative effects. We left this for future studies.
%\subsection{D2 brane descrption}
%One possible origin of this mismatch may come from the existence of
%an interpolating function between strong and weak coupling. One may
%propose that
% \be
 %\langle W(circle)\rangle \sim exp(2\pi f(\lambda),
 %\ee
%with $f(\lambda)$ approaches to $\sqrt{\lambda/2}$ at strong
%coupling and to $\lambda$ at weak
%coupling\cite{Gaiotto:2008cg,Nishioka:2008gz,Grignani:2008is,
%ChenWu2008}. However, even taking into account of this fact, there
%is still a numerical mismatch, which suggests that the interpolating
%function could be not universal.
%\subsection{D6 brane descrption}

\section{Conclusion and Discussions}

In this paper, we studied the supersymmetric Wilson loops in ABJM
theory from both sides of the $AdS_4/CFT_3$ correspondence. In the
field theory side, we constructed $1/6$-BPS Wilson loop operators
and perform the perturbative computations of their VEVs at the first
two orders. In the string theory side, we found the fundamental
string solution which could be dual to half-BPS circular Wilson
loop.

There are a lot of interesting open questions we left for further
studies. Here we list some of them. One natural question is to
search for other BPS Wilson loops, especially the half-BPS one which
seems to us not easy to construct. Even for the $1/6$-BPS Wilson
loops in this paper, our understanding is quite limited. Recently a
class of Wilson loops operators are constructed in ABJM theory based
on M-theory considerations \cite{Berenstein} following the line in
\cite{Maldacena98}.  It is interesting to check if the $1/6$-BPS
Wilson loops belong to that class of operators and search for new
supersymmetric Wilson loops in that class.

We also notice that Berkovits' pure string formalism
\cite{Berkovits00, Berkovits02} has been used in the string theory
dual of ABJM theory \cite{Fre:2008qc, Bonelli:2008us,
D'Auria:2008cw}. Based on these works, maybe one can study the
Wilson loops in ABJM theory using the ideas in \cite{Bonelli:2008rv}
for the $AdS_5$ case.

The fundamental string dual to half-BPS operator is almost the same
as the one in $AdS_5$ case. However the string solution dual to the
$1/6$-BPS operators may also be  embedded in the $CP^3$ part of the
background in a non-trivial manner as the string solution in
$AdS_5\times S^5$ dual to $1/4$-BPS Wilson loop in super Yang-Mills
theory \cite{Drukker:2006ga}. It is interesting to explicitly find
this solution. The probe brane dual to Wilson loops in higher
dimensional representations need further studies as well. Another
interesting question is to find $M2$-brane in $AdS_4\times S^7/Z_k$
dual to these Wilson loops.

Some membrane solutions in $AdS_4\times S^7$ is found in
\cite{Lunin, Chen}. By generalizing them to membrane solutions
$AdS_4 \times S^7/Z_k$ and using the reduction from M-theory to IIA
theory, it is reasonable to get some fundamental string solution in
$AdS_4\times CP^3$. The field theory dual of these string solutions
demands further studies.

\section*{Acknowledgments}
The work was partially supported by NSFC Grant No.10535060,10775002
and NKBRPC (No. 2006CB805905). JB would like to thank Riccardo
Argurio, Matteo Bertolini, Giulio Bonelli, Nadav Drukker, Edi Gava,
Chethan Gowdigere, Qing-Guo Huang, Kimyeong Lee, Juan Maldacena, Jan
Plefka, Houman Safaai, Takao Suyama, Mahdi Torabian and Ho-Ung Yee
for helpful discussions. We also thank Nadav Drukker and Jan Plefka
for correspondence prior to submission. The Feynman diagrams are
drawn using Axodraw \cite{axodraw}.

\appendix

\section{The Feynman rules}

 \begin{figure}[ht!]
    \epsfxsize=50mm%
    \hfill\epsfbox{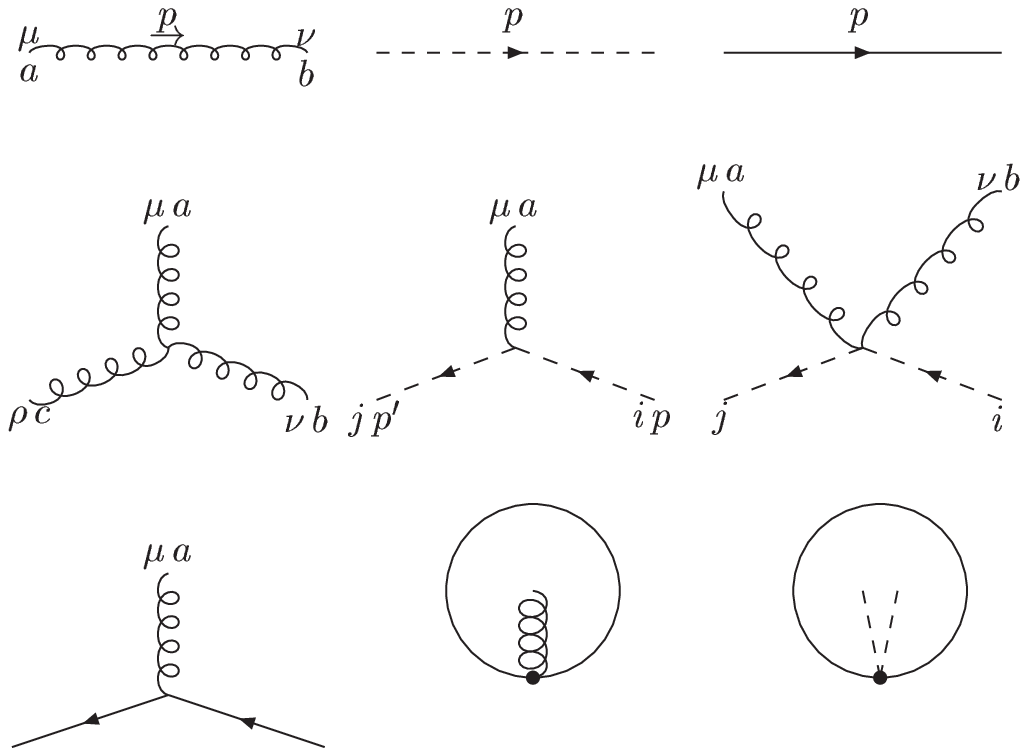}\hfill~\\
    \caption{Some needed Feynman rules}
    %\label{figLT}
   \end{figure}

In this appendix, we list some of the Feynman rules needed in our
perturbative calculation in ABJM theory. The diagrams are put in
Fig.~6, and the corresponding expressions are the following:
\begin{enumerate}
 \item  Gluon propagator
 \be\pm\d^{ab}\frac{4\pi}k\epsilon^{\m\n\lam}p_\lam\frac1{p^2} \ee
 in the Landau gauge, where the up/down sign is corresponding to
 the first/sector factor of the gauge group.
 \item Scalar propagator:
$\frac{i}{p^2}$.
 \item Fermion propagator:
 $\frac{ip_\mu\g^\m}{p^2}$.
 \item Three-gluon vertex:
$\pm \frac{k}{4\pi}f^{abc}\epsilon^{\m\n\lam}$. The up/down sign is
corresponding to the first/sector factor of the gauge group.
 \item Gluon-scalar-scalar vertex:
$-i(p+p^\prime)^\m(T^a)_{ij}$.
\item gluon-gluon-scalar-scalar vertex:
 $=i\eta^{\m\n}(\{T_a, T_b\})_{ij}$
\item Gluon-fermion-fermion vertex:
$i \g^\m(T^a)$.
\item One gluon emitted the Wilson loop:
$i$.
\item Two scalars emitted the Wilson loop:
$\pm i\frac{2\pi}k$ for $A$'s/$B$'s.
\end{enumerate}
The last two vertices are from the Wilson loop operator (we omit the
color factors in the expressions of these two vertices). We don't
give the explicit express from the six-scalar vertex and
scalar-scalar-fermion-fermion vertex. We also omit the ghost
propagator and gluon-ghost-ghost vertex.

\ed